 \documentclass[final,3p,times]{elsarticle}
\usepackage{amssymb}
\usepackage{lineno}
\usepackage{color}
\journal{Journal Name}

\newtheorem{thm}{Theorem}

\newdefinition{rmk}{Remark}
\newtheorem{prop}{Proposition}
\newtheorem{coro}{Corollary}
\newproof{pf}{Proof}
\newproof{pot}{Proof of Theorem \ref{thm2}}

\begin{document}

\begin{frontmatter}

\title{$\beta$-Divergence loss for the kernel density estimation with bias reduced}

\author[mysecondaryaddress]{Hamza Dhaker\corref{mycorrespondingauthor}}
\cortext[mycorrespondingauthor]{Corresponding author}
\ead{hamza.dhaker@umoncton.ca}
\author[mymainaddress]{El Hadji Deme}
\author[mymainaddress]{and Youssou Ciss}
\address[mysecondaryaddress]{D\'{e}partement de math\'{e}matiques et statistique,Universit\'{e} de Moncton, NB, Canada }
\address[mymainaddress]{LERSTAD,UFR SAT, Universite Gaston Berger, Saint-Louis, Senegal}

\begin{abstract}
Allthough nonparametric kernel density estimation with bias reduce is nowadays a standard technique in explorative data-analysis, there is still a big dispute on how to assess the quality of the estimate and which choice of bandwidth is optimal. This article examines the most important bandwidth selection methods for kernel density estimation with bias reduce, in particular, normal reference, least squares cross-validation, biased crossvalidation and $\beta$-Divergence loss. Methods are described and expressions are presented. We will compare these various bandwidth selector on simulated data. As an example of real data,  we will use econometric data sets CO2 per capita in example 1 and the second data set consists of 107 eruption lengths in minutes for the Old Faithful geyser in Yellowstone National Park, USA.

\end{abstract}

\end{frontmatter}

%\linenumbers

\section{Introduction}
\label{Intro}
Selecting an appropriate bandwidth for a kernel density estimator is of crucial importance, and the purpose of the estimation may be an influential factor in the selection method. In many situations, it is sufficient to subjectively choose the smoothing parameter by looking at the density estimates produced by a range of bandwidths. A good overview on kernel density estimators is supplied by Silverman \cite{sil}; Scott \cite{sco}; Mugdadi and Ahmad \cite{mug}.
\\

Throughout this article, we use the following notation. Let $X_1,..., X_n$, of size $n$ from a density $f$ where the Parzen \cite{par} kernel estimator $f_{n}$ is defined by
$$ f_{n}(x)=\frac{1}{nh}\sum_{i=1}^{n}{\cal K}(\frac{x-X_{i}}{h}) $$
In general, ${\cal K}$ is the kernel function (e.g. normal density function) and $h$ is the bandwidth. A large body of research is devoted to choosing $h$, essentially the amount of smoothing to apply. Usually, smoothing parameters can be chosen via cross validation or by minimizing a measure of error.
\\

An important recent paper in this area is Xie and Wu \cite{xie}. Xie and Wu studied  an estimator which reduces the bias, the performance of which in both theory and simulations proved to be clearly superior to other methods currently popular in the literature. Xie and Wu \cite{xie} presented an estimator which reduces the bias, defined by:
\begin{eqnarray}
\label{fn}
\hat{f}_{n}(x) &=& f_{n}(x)-\widehat{Bias}(f_{n}(x)) \\ &=&  f_{n}(x)-\frac{h^{2}}{2}f^{''}_{n}(x)\int t^{2}{\cal K}(t)dt
\end{eqnarray}

the bandwidth is the most dominant parameter in the kernel density estimator. This parameter controls the amount of smoothing, and is analogous to the bandwidth in a histogram.
Even though the kernel estimator depends on the kernel and the bandwidth in a rather complicated way, a graphical representation clearly illustrates the difference in importance between these two parameters, see figure 2.3 and 2.6a in Wand and Jones \cite{wan}. To explore the most relevant bandwidth selection methods in density estimation for complete data see the reviews of Turlach \cite{tur}, Cao et al. \cite{cao}, Jones et al. \cite{jon} or Heidenreich et al. (2013), Mammen et al. (\cite{mam1} and \cite{mam2}), and the recent work  on $\beta$-Divergence for Bandwidth Selection by Dhaker and al. \cite{ham}.
\\
Our aim in this paper is to propose and compare several bandwidth selection procedures for the kernel density estimators introduced by Xie and Wu \cite{xie}. The procedures we study are bandwidth selector based on the criterion of $\beta$-divergence with different beta values. A simulation study is then carried out to assess the finite sample behavior of these bandwidth selectors. \\
The remainder of the paper is organised as follows. In Section 2, we state our main results: presents the method proposed for bandwidth selector based ${\cal D}_{\beta}$. Section 3 estimation of the optimal bandwidth
Section 4 is devoted to our simulation results. Section 5 applies the methods to real data. We conclude the paper in Section 6,

\section{Bandwidth selection based $\beta$-divergence}
The $\beta$-divergence [Cichocki \cite{cic}, Basu \cite{bas} and Eguchi \cite{egu}] is a general framework of similarity measures induced 
from various statistical models, such as Poisson, Gamma, Gaussian, Inverse Gaussian and compound Poisson distribution. For 
the connection between the $\beta$-divergence and the various statistical distributions, see Jorgensen \cite{jor}.
Beta divergence was proposed in Basu \cite{bas} and Minami and Eguchi \cite{min} and is defined as dissimilarity between the density function and its estimator
\begin{eqnarray*}
D_{\beta}(\hat{f}_{n}(x) ,f(x))=\frac{1}{\beta}\int\hat{f}_{n}^{\beta}(x)dx -\frac{1}{\beta -1}\int\hat{f}_{n}^{\beta -1}(x)f(x)dx +\frac{1}{\beta (\beta -1)}\int\ f^{\beta}(x)dx.
\end{eqnarray*}
In the case $\beta = 2$,
$$ 2D_{2}(\hat{f}_{n}(x) ,f(x))= ISE(\hat{f}_{n}(x))= \int (\hat{f}_{n}(x) -f(x))^{2}dx.$$

The following theorem allows us to give the analytical value of bandwidth which minimizes the mean $D_{\beta}(\hat{f}_{n}(x) ,f(x))$.  
\begin{thm}
\label{theo1}
Let the following conditions on $f$ be satisfied:
\begin{itemize}
\item[(F1)] $f$ is compactly supported on $I$.
\item[(F2)] $f$ is four times continuously differentiable on $I$.
%\item[(F3)] $ \displaystyle \lim_{x\longrightarrow +\inf I }f^{(i)}(x)= \displaystyle \lim_{x\longrightarrow -\sup I }f^{(i)}(x) $ , $1\leq j \leq 3$. 
\item[(F3)] $\int_{I}f^{(4)}(x)^{2}f(x)^{\beta -2}dx <\infty $.
\end{itemize}
the bandwidth $h_{\mathbb{E}D_{\beta}}$ that minimizes the mean $\beta$-divergence between a kernel estimator $\widehat{f}_{h}$ and density $f$ is:
\begin{eqnarray}
\label{hEDbth}
h_{\beta}=h_{\mathbb{E}{\cal D}_{\beta}}=\left\lbrace  72\frac{\int {\cal K}(t)^{2}dt\int_{I} f(x)^{\beta -1}dx }{  \left[ \int t^{4}{\cal K}(t)dt \right]^{2} \int_{I} f(x)^{\beta -2}\left( f^{(4)}(x)\right)^{2}dx } \right\rbrace ^{1/9}n^{-1/9}
\end{eqnarray}

\end{thm}

\begin{prop}
\label{prop1}
Under $(F1)-(F3)$ we have
\begin{eqnarray}
\label{estEDb}
 \mathbb{E} {\cal D}_{\beta}(\hat{f}_{n}(x) ,f(x)) = \frac{1}{2}\left[  \frac{h^{8}}{576}\left(\int_{I}t^{4}{\cal K}(t)dt\right)^{2}\int f^{\beta-2}(x)\left( f^{(4)}(x)\right)^{2}dx +\frac{1}{nh}\int_{I}{\cal K}^{2}(t)dt \int f^{\beta -1}(x)dx  \right] +O_{p}(n^{-c}) +O(h^{6}) 
\end{eqnarray}
and
\begin{eqnarray}
\label{estEDb}
 A\mathbb{E} {\cal D}_{\beta}(\hat{f}_{n}(x) ,f(x)) = \frac{1}{2}\left[  \frac{h^{8}}{576}\left(\int_{I}t^{4}{\cal K}(t)dt\right)^{2}\int f^{\beta-2}(x)\left( f^{(4)}(x)\right)^{2}dx +\frac{1}{nh}\int_{I}{\cal K}^{2}(t)dt \int f^{\beta -1}(x)dx  \right]
\end{eqnarray}
where $0<c<\frac{1}{8}$
\end{prop}

\begin{coro}
Assuming that the assumptions in Theorem \ref{theo1} hold with $\beta =2$. Then, we have
$$\mathbb{E}{\cal D}_{2}(\widehat{f}_{h},f)= \frac{1}{2} MISE(\widehat{f}_{h},f)$$
$$A\mathbb{E}{\cal D}_{2}(\widehat{f}_{h},f)= \frac{1}{2} AMISE(\widehat{f}_{h},f)$$
in that case
\begin{eqnarray}
\label{hMISE}
h_{AMISE(\widehat{f}_{h},f)}=\left\lbrace  \frac{9}{2}\frac{ R({\cal K})}{  \left( \mu_{4}({\cal K}) \right)^{2} R(f^{(4)}) } \right\rbrace ^{1/9}n^{-1/9}
\end{eqnarray}

$R(g)=\int g(t)^{2}dt \qquad $ and $\qquad \mu_{4}({\cal K})=\int x^{4}{\cal K}(x)dx $

\end{coro}

\begin{pf}

\begin{eqnarray*}
\widehat{f}_{n}^{\beta}(x)=\left( f_{n}(x) - \widehat{Bias}(\widehat{f}(x)\right) ^{\beta}
\end{eqnarray*}

With a random variable $\xi =O_{p}(1)$ whose expectation is $0$ and variance $1$, we can write $f_{n}(x)$ as (see Kanazawa \cite{kan})

\begin{eqnarray}
f_{n}(x)=f(x)\left[ 1+\frac{h^{2}}{2}\frac{f^{(2)}(x)}{f(x)}\int_{I}t^{2}{\cal K}(t)dt +\frac{h^{4}}{24}\frac{f^{(4)}(x)}{f(x)}\int_{I}t^{4}{\cal K}(t)dt +O(h^{6}) + \left\lbrace \frac{\int_{I}{\cal K}(t)^{2}dt}{nhf(x)}  \right\rbrace ^{1/2}\xi + O_{p}(n^{-1/2}) \right]
\end{eqnarray}

%\begin{eqnarray}
%E(f_{n}(x))=f(x)\left[ 1+\frac{h^{2}}{2}\frac{f^{(2)}(x)}{f(x)}\int_{I}t^{2}{\cal K}(t)dt +\frac{h^{4}}{24}\frac{f^{(4)}(x)}{f(x)}\int_{I}t^{4}{\cal K}(t)dt +O(h^{5})  \right],
%\end{eqnarray}

Using the result of the corollary 2.6 \cite{eug}
$$\lim_{n\rightarrow \infty} \sup_{x} n^{c}\vert f_{n}^{(r)}(x)-f^{(r)}(x)\vert =0 \quad with \quad 0 < c < \frac{1}{2r+4} $$,
we have

\begin{eqnarray*}
\widehat{f}_{n}(x)&=& f_{n}(x) - \widehat{Bias}(f_{n}(x))=f_{n}(x) -\frac{h^{2}}{2}f_{n}^{(2)}\int_{I} t^{2}K(t)dt= f_{n}(x) -\frac{h^{2}}{2}f^{(2)}\int_{I} t^{2}K(t)dt + O(n^{-c})\\ &=& f(x)\left[ 1+\frac{h^{2}}{2}\frac{f^{(2)}(x)}{f(x)}\int_{I}t^{2}{\cal K}(t)dt +\frac{h^{4}}{24}\frac{f^{(4)}(x)}{f(x)}\int_{I}t^{4}{\cal K}(t)dt +O(h^{6}) + \left\lbrace \frac{\int_{I}{\cal K}(t)^{2}dt}{nhf(x)}  \right\rbrace ^{1/2}\xi + O_{p}(n^{-1/2}) \right] \\ && -\frac{h^{2}}{2}f^{(2)}\int_{I} t^{2}K(t)dt+O(n^{-c}) \\ &=& f(x)\left[ 1+\frac{h^{4}}{24}\frac{f^{(4)}(x)}{f(x)}\int_{I}t^{4}{\cal K}(t)dt +O(h^{6}) + \left\lbrace \frac{\int_{I}{\cal K}(t)^{2}dt}{nhf(x)}  \right\rbrace ^{1/2}\xi + O_{p}(n^{-1/2}) +O(n^{-c})\right]
\end{eqnarray*}

Where the $ O(h^{6}) $ terms depend upon $x$. Using $(1+z)^{\beta}=1 + \beta z +\frac{\beta (\beta -1)}{2}z^{2} + O(z^{3})$
\begin{eqnarray*}
\widehat{f}_{n}^{\beta}(x)  &=& f(x)^{\beta}\left[ 1+\frac{h^{4}}{24}\frac{f^{(4)}(x)}{f(x)}\int_{I}t^{4}{\cal K}(t)dt +O(h^{6}) + \left\lbrace \frac{\int_{I}{\cal K}(t)^{2}dt}{nhf(x)}  \right\rbrace ^{1/2}\xi + O_{p}(n^{-1/2}) +O(n^{-c}) \right]^{\beta} \\ &=& f(x)^{\beta} \left[ 1+\beta\left( \frac{h^{4}}{24}\frac{f^{(4)}(x)}{f(x)}\int_{I}t^{4}{\cal K}(t)dt+\left\lbrace \frac{\int_{I}{\cal K}(t)^{2}dt}{nhf(x)}  \right\rbrace ^{1/2}\xi\right)+ \frac{\beta (\beta-1)}{2}\left( \frac{h^{8}}{576}\frac{(f^{(4)}(x))^{2}}{f^{2}(x)}(\int_{I}t^{4}{\cal K}(t)dt)^{2}+  \frac{\int_{I}{\cal K}(t)^{2}dt}{nhf(x)} \xi^{2}\right)\right. \\  && + \left. O_{p}(n^{-c}) +O(h^{6})  \right],
\end{eqnarray*}
and
\begin{eqnarray*}
\widehat{f}_{n}^{\beta -1}(x)  &=& f(x)^{\beta -1}\left[ 1+\frac{h^{4}}{24}\frac{f^{(4)}(x)}{f(x)}\int_{I}t^{4}{\cal K}(t)dt +O(h^{6}) + \left\lbrace \frac{\int_{I}{\cal K}(t)^{2}dt}{nhf(x)}  \right\rbrace ^{1/2}\xi + O_{p}(n^{-1/2}) +O(n^{-c}) \right]^{\beta -1} \\ &=& f(x)^{\beta -1} [ 1+(\beta -1)\left( \frac{h^{4}}{24}\frac{f^{(4)}(x)}{f(x)}\int_{I}t^{4}{\cal K}(t)dt+\left\lbrace \frac{\int_{I}{\cal K}(t)^{2}dt}{nhf(x)}  \right\rbrace ^{1/2}\xi\right) \\  &&+ \frac{(\beta -1)(\beta -2)}{2}\left( \frac{h^{8}}{576}\frac{(f^{(4)}(x))^{2}}{f^{2}(x)}(\int_{I}t^{4}{\cal K}(t)dt)^{2}+  \frac{\int_{I}{\cal K}(t)^{2}dt}{nhf(x)} \xi^{2}\right) \\  && +O_{p}(n^{-c}) +O(h^{6})  ]
\end{eqnarray*}
 
\begin{eqnarray*}
{\cal D}_{\beta}(\hat{f}_{n}(x) ,f(x)) &=& \frac{1}{\beta}\int\hat{f}_{n}^{\beta}(x)dx -\frac{1}{\beta -1}\int\hat{f}_{n}^{\beta -1}(x)f(x)dx +\frac{1}{\beta (\beta -1)}\int\ f^{\beta}(x)dx \\ &=& \frac{1}{\beta}\int f(x)^{\beta} [ 1+\beta\left( \frac{h^{4}}{24}\frac{f^{(4)}(x)}{f(x)}\int_{I}t^{4}{\cal K}(t)dt+\left\lbrace \frac{\int_{I}{\cal K}(t)^{2}dt}{nhf(x)}  \right\rbrace ^{1/2}\xi\right)\\  && + \frac{\beta (\beta-1)}{2}\left( \frac{h^{8}}{576}\frac{(f^{(4)}(x))^{2}}{f^{2}(x)}(\int_{I}t^{4}{\cal K}(t)dt)^{2}+  \frac{\int_{I}{\cal K}(t)^{2}dt}{nhf(x)} \xi^{2}\right) \\  && +O_{p}(n^{-c}) +O(h^{6})  ] dx \\ && -\frac{1}{\beta -1}\int f(x)^{\beta} [ 1+(\beta -1)\left( \frac{h^{4}}{24}\frac{f^{(4)}(x)}{f(x)}\int_{I}t^{4}{\cal K}(t)dt+\left\lbrace \frac{\int_{I}{\cal K}(t)^{2}dt}{nhf(x)}  \right\rbrace ^{1/2}\xi\right) \\  &&+ \frac{(\beta -1)(\beta -2)}{2}\left( \frac{h^{8}}{576}\frac{(f^{(4)}(x))^{2}}{f^{2}(x)}(\int_{I}t^{4}{\cal K}(t)dt)^{2}+  \frac{\int_{I}{\cal K}(t)^{2}dt}{nhf(x)} \xi^{2}\right) \\  && +O_{p}(n^{-c}) +O(h^{6})  ]dx \\ && + \frac{1}{\beta (\beta -1)}\int\ f^{\beta}(x)dx \\  &=&
 \frac{1}{\beta}\int f(x)^{\beta} [ \frac{\beta (\beta-1)}{2}\left( \frac{h^{8}}{576}\frac{(f^{(4)}(x))^{2}}{f^{2}(x)}(\int_{I}t^{4}{\cal K}(t)dt)^{2}+  \frac{\int_{I}{\cal K}(t)^{2}dt}{nhf(x)} \xi^{2}\right) +O_{p}(n^{-c}) +O(h^{6})  ] dx \\ && -\frac{1}{\beta -1}\int f(x)^{\beta} [ \frac{(\beta -1)(\beta -2)}{2}\left( \frac{h^{8}}{576}\frac{(f^{(4)}(x))^{2}}{f^{2}(x)}(\int_{I}t^{4}{\cal K}(t)dt)^{2}+  \frac{\int_{I}{\cal K}(t)^{2}dt}{nhf(x)} \xi^{2}\right) + O_{p}(n^{-c}) +O(h^{6})  ]dx \\  &=&
 \int f(x)^{\beta} [ (\frac{\beta-1}{2}-\frac{\beta-2}{2})\left( \frac{h^{8}}{576}\frac{(f^{(4)}(x))^{2}}{f^{2}(x)}(\int_{I}t^{4}{\cal K}(t)dt)^{2}+  \frac{\int_{I}{\cal K}(t)^{2}dt}{nhf(x)} \xi^{2}\right) +O_{p}(n^{-c}) +O(h^{6})  ] dx  \\  &=&
 \frac{1}{2}\left[  \frac{h^{8}}{576}(\int_{I}t^{4}{\cal K}(t)dt)^{2}\int f^{\beta-2}(x)\left( f^{(4)}\right)^{2}(x)dx +\frac{1}{nh}\int_{I}{\cal K}^{2}(t)dt \int f^{\beta -1}(x)dx \xi^{2} \right] +O_{p}(n^{-c}) +O(h^{6}) 
\end{eqnarray*}

\begin{eqnarray*}
 \mathbb{E} {\cal D}_{\beta}(\hat{f}_{n}(x) ,f(x)) = \frac{1}{2}\left[  \frac{h^{8}}{576}\left(\int_{I}t^{4}{\cal K}(t)dt\right)^{2}\int f^{\beta-2}(x)\left( f^{(4)}\right)^{2}(x)dx +\frac{1}{nh}\int_{I}{\cal K}^{2}(t)dt \int f^{\beta -1}(x)dx  \right] +O_{p}(n^{-c}) +O(h^{6}) 
\end{eqnarray*}

\end{pf}

\newpage

\section{The choice of the bandwidth $h$}
\label{choi}
In this section, we describe bandwidth selection methods for the density estimator defined in (1). These methods consist of adaptations of common automatic selectors for kernel density estimation. We propose two selection methods a Normal reference and the cross-validation method. The Normal reference bandwidth is based on estimating the infeasible optimal expression (\ref{hMISE}), in which the unknown element is $R(f^{(4)})$.

\subsection{Rule-of-thumb for bandwidth selection}
This method is based on the rule-of-thumb, Silverman \cite{sil}, for complete data.The idea
is to assume that the underlying distribution is normal, $ {\cal N}(\mu ,\sigma )$, and in this situation

$$f(x)=\frac{1}{\sigma\sqrt{2\pi}}e^{-\frac{1}{2}(\frac{x-m}{\sigma})^{2}},$$
so 
$$f^{(4)}(x)= \frac{1}{\sigma^{5}\sqrt{2\pi}}e^{-\frac{1}{2}(\frac{x-m}{\sigma})^{2}}\left( 3 -6\left(\frac{x-m}{\sigma}\right)^{2} +\left(\frac{x-m}{\sigma}\right)^{4} \right), $$

$$\left( f^{(4)}(x)\right)^{2}= \frac{1}{\sigma^{10} 2\pi}e^{-(\frac{x-m}{\sigma})^{2}}\left( 9-36\left(\frac{x-m}{\sigma}\right)^{2} + 30\left(\frac{x-m}{\sigma}\right)^{4} +18\left(\frac{x-m}{\sigma}\right)^{6}+\left(\frac{x-m}{\sigma}\right)^{8} \right), $$

\begin{eqnarray*} 
 \int f^{\beta -2}(x) \left(f^{(4)}(x)\right)^{2}dx = \frac{1}{\sigma^{\beta +7}\sqrt{\beta}(2\pi)^{\frac{\beta -2}{2}} }\left( \frac{9\beta^{4}-36\beta^{3}+90\beta^{2}+270\beta +105}{\beta^{4}} \right).
\end{eqnarray*}

and

$$ \int f^{\beta -1}(x)dx = \frac{1}{\sqrt{\beta -1}(2\pi)^{\frac{\beta -2}{2}}} $$
In that case the asymptotically optimal bandwidth $h_{\beta}$ in Equation (\ref{hEDbth}) becomes the normal reference bandwidth.

\begin{eqnarray*}
h_{\beta}=h_{\mathbb{E}{\cal D}_{\beta}} &=&\left\lbrace  72\frac{R({\cal K})\int_{I} f(x)^{\beta -1}dx }{ \mu_{4}({\cal K})^{2} \int_{I} f(x)^{\beta -2}\left( f^{(4)}(x)\right)^{2}dx } \right\rbrace ^{1/9}n^{-1/9} \\ &=&
 \left\lbrace  72\frac{R({\cal K})}{\sqrt{\beta -1}(2\pi)^{\frac{\beta -2}{2}} \mu_{4}({\cal K})^{2} \frac{1}{\sigma^{\beta +7}\sqrt{\beta}(2\pi)^{\frac{\beta -2}{2}} }\left( \frac{9\beta^{4}-36\beta^{3}+90\beta^{2}+270\beta +105}{\beta^{4}} \right) } \right\rbrace ^{1/9}n^{-1/9}
\end{eqnarray*} 

with $\sigma$ being the standard deviation of $f$.\\
For the Gaussian kernel, $\mu_{4}({\cal K}) =3$ and $R({\cal K})=(4\pi)^{-\frac{1}{2}}$ so that
\begin{eqnarray}
h_{NR_{\beta}}=\left\lbrace \sqrt{\frac{2}{\pi}}\frac{4\beta^{4}}{9\beta^{4}-36\beta^{3}+90\beta^{2}+27\beta +105}\frac{1}{n} \right\rbrace ^{1/9}\sigma
\end{eqnarray}
for $\beta =2$
\begin{eqnarray}
h_{NR_{2}}=\left\lbrace \sqrt{\frac{16}{861}\frac{2}{\pi}}\frac{1}{n} \right\rbrace ^{1/9}\sigma .
\end{eqnarray}
The standard deviation $\sigma$ can be estimated by the sample standard deviation $s$ or by the standardized interquartile range $IQR/1.34$ for robustness against outliers $(1.34 = \Phi^{-1}(3/4)- \Phi^{-1}(1/4))$, but a better rule of thumb is (e.g., Silverman, 1986, pp. 45–47; Härdle, 1991, p. 91).

\begin{eqnarray}
\widehat{h}_{NR_{2}}=\left\lbrace \sqrt{\frac{16}{861}\frac{2}{\pi}}\frac{1}{n} \right\rbrace ^{1/9}\widehat{\sigma},
\end{eqnarray}
with $\widehat{\sigma} = \min (s, IQR/1.34)$

\subsection{Cross-validation}
The method previously defined is based on minimising estimations of the $MISE$, more precisely of the $AMISE$. This 
procedure relies on the minimisation of the $ISE$ (integrated squared error), the methodology is the same as in Rudemo 
\cite{rud} and Bowman \cite{bow} applied to (\ref{fn}).

Let write: $$ 
D_{\beta}(\hat{f}_{n}(x) ,f(x))=\frac{1}{\beta}\int\hat{f}_{n}^{\beta}(x)dx -\frac{1}{\beta -1}\int\hat{f}_{n}^{\beta -1}(x)f(x)dx +\frac{1}{\beta (\beta -1)}\int\ f^{\beta}(x)dx ,$$
Note that $\frac{1}{\beta (\beta -1)}\int\ f^{\beta}(x)dx$ dz does not depend on $h$, so the minimisation of the ISE is equivalent to minimise the following function:
\begin{eqnarray*}
 L(h)&=& D_{\beta}(\hat{f}_{n}(x) ,f(x))- \frac{1}{\beta (\beta -1)}\int\ f^{\beta}(x)dx ,\\ &=&
 \frac{1}{\beta}\int\hat{f}_{n}^{\beta}(x)dx -\frac{1}{\beta -1}\int\hat{f}_{n}^{\beta -1}(x)f(x)dx ,\\ &=&
 \frac{1}{\beta}\int\hat{f}_{n}^{\beta}(x)dx -\frac{1}{\beta -1}\mathbb{E}\left( \hat{f}_{n}^{\beta -1}(x)\right).
\end{eqnarray*}
The principle of the least squares cross-validation method is to find an estimate of $L(h)$ from the data and minimize it over $h$. Consider the estimator
$$LSCV(h)=  \frac{1}{\beta}\int\hat{f}_{n}^{\beta}(x)dx -\frac{2}{n}\frac{1}{\beta -1}\sum_{i=1}^{n}  \hat{f}_{h(i)}^{\beta -1}(X_{i}), $$
with $\hat{f}_{h(i)}^{\beta -1}(X_{i})=\frac{1}{h(n-1)}\sum_{j\neq i}^{n}{K}\left(\frac{X_{i}-X_{j}}{h} \right).$
\section{Simulation}
In this section we evaluate the performance of the bandwidth selection procedures presented in Section 2. To this goal we have carried out a simulation study including rule-of-thumb ($h_{NR}$), cross-validation bandwidth ($h_{LSCV}$) and $h_{\beta}$ for minimizing criterion $\beta$-divergence (with $\beta = 1.5,  1.1 and 1.9$).
\\

We consider various sets of experiments in which data are generated from the mixture of a Normal ${\cal N}(0, 1)$ and Normal ${\cal N}(\mu, \sigma)$ distributions. Hence, the DGP (Data Generating Process) is generated
from $m(\mu , \sigma^{2})$ with the density
\begin{eqnarray}
\label{mpi}
m(\mu , \sigma^{2}) = 0.5{\cal N}(0, 1) + 0.5{\cal N}(\mu, \sigma)
\end{eqnarray}

where $\mu =0, 1, 5$ and $\sigma =1, 0.5, 0.1$. One thousand Monte Carlo samples of size n are generated from the normal mixture model in Equation (\ref{mpi}) for each combination of $n = 50, 200, 700$.
The results of our different sets of experiments are presented in Tables 1-3. \\
Table 1 give the exhibits simulated relative efficiency $RE(\widehat{h})=MISE(\widehat{f}_{\widehat{h}_{MISE}})/MISE(\widehat{f}_{\widehat{h}})$ of the kernel estimator, using bandwidths $\widehat{h}_{NR}$, $\widehat{h}_{LSCV}$ and $\widehat{h}_{\beta}$, it is lower than 1, because the optimal bandwidth $h_{MISE}$ minimize $MISE$. Each bandwidth, mean $\mathbb{E}(\widehat{h})$ and mean relation error  $\mathbb{E}(\widehat{h}/h_{MISE}-1)$ are obtained, these values are given by respectively,
Tables 2 and 3.
\\
\begin{itemize}
\item[1-] For all situations, each relative efficiency $RE(\widehat{h})< 1$ because the optimal
bandwidth $h_{MISE}$ minimizes the $MISE$.
\item[2-] The normal reference bandwidth hNR performswell if the true density is not very far from normal, such as the cases of $(\mu, \sigma) = (0, 1), (0, 0.5), (1, 1)$, and $(1, 0.5)$. Otherwise, it usually has the smallest $RE(\widehat{h})$ and largest $E(\widehat{h})$, tending to oversmooth its kernel density estimate the most.
\item[2-] We have to remark that in table 1, the LSCV bandwidth hlscv needs a large sample size in order to be competitive. Note also that in Table 2, it is seen that $\mathbb{E}(\widehat{h}_{LSCV})$ is close to the optimal $\widehat{h}_{MISE}$, but the corresponding $\mathbb{E}(\widehat{h}_{LSCV}/\widehat{h}_{MISE})$ is large, which means that the bias of $\widehat{h}_{LSCV}$ is small but its variation is large in Table 3.
\item[3-] The  bandwidth $\widehat{h}_{\beta}$ seems to be the best existing bandwidth selectors. In most situations, it is indeed one  of the best bandwidth selectors, However, it behaves very poorly for small $\sigma$ (the true density curve is sharp).
\end{itemize}

%%%%%%%%%%%%%%%%%%%%%%%%%%%%%%%%%%%%%%%%%%%%%%%%%%%%%%%%%

\begin{table}[h]
\begin{center}
\begin{small}
\caption{$RE \left(\widehat{h} \right) $ for normal mixture $f(x)=0.5 \phi (x)+0.5\phi_{\sigma}(x-\mu)$ \label{tab-id_table1}}
\begin{tabular}{ccccccccc}
  \hline 
  {\small $n$} & {\small $\widehat{h}_{NR}$} & {\small $\widehat{h}_{LSCV}$ }  & {\small $\widehat{h}_{{\cal D}_{1.1}CV}$} & {\small $\widehat{h}_{{\cal D}_{1.5}CV}$} & {\small $\widehat{h}_{{\cal D}_{1.9}CV}$}  \\
 \hline 
  & &  {\scriptsize $\mu =0$}& {\scriptsize $\sigma=1$ }& \\
 {\scriptsize 50} & {\scriptsize 0.934} &  {\scriptsize 0.953} & {\scriptsize 0.853} & {\scriptsize 0.723} & {\scriptsize 0.703}  \\ 
 {\scriptsize 200} & {\scriptsize 0.945} &{\scriptsize 0.925}  & {\scriptsize 0.955} & {\scriptsize 0.931} & {\scriptsize 0.903}   \\   
 {\scriptsize 700} & {\scriptsize 0.990} &{\scriptsize 0.945}  &{\scriptsize 0.982} & {\scriptsize 0.987} & {\scriptsize 0.952} \\ 
 
  & &  {\scriptsize $\mu =0$} & {\scriptsize $\sigma=0.5$}  & \\
 {\scriptsize 50} &{\scriptsize 0.870} &{\scriptsize 0.837}  &{\scriptsize 0.867} & {\scriptsize 0.890} & {\scriptsize 0.905}   \\ 
 {\scriptsize 200} & {\scriptsize 0.937} & {\scriptsize 0.880} &{\scriptsize 0.897} & {\scriptsize 0.954} & {\scriptsize 0.932} \\  
 {\scriptsize 700} & {\scriptsize 0.964} & {\scriptsize 0.930}  &{\scriptsize 0.929} & {\scriptsize 0.842} & {\scriptsize 0.858}   \\ 

  & & {\scriptsize $\mu =0$}& {\scriptsize $\sigma=0.1$} & \\
 {\scriptsize 50} &{\scriptsize 0.584} &{\scriptsize 0.767} &{\scriptsize 0.634} & {\scriptsize 0.631} & {\scriptsize 0.623}   \\ 
 {\scriptsize 200} &{\scriptsize 0.553} &{\scriptsize 0.892} &{\scriptsize 0.625} & {\scriptsize 0.879} & {\scriptsize 0.721} \\ 
 {\scriptsize 700} &{\scriptsize 0.529} &{\scriptsize 0.946} &{\scriptsize 0.612} & {\scriptsize 0.877} & {\scriptsize 0.813}   \\  
  
  & & {\scriptsize $\mu =1$}& {\scriptsize $\sigma=1$} & \\ 
 {\scriptsize 50}&{\scriptsize 0.864}&{\scriptsize 0.899} &{\scriptsize 0.904} & {\scriptsize 0.853} & {\scriptsize 0.876}  \\  
 {\scriptsize 200} &{\scriptsize 0.938} &{\scriptsize 0.928} &{\scriptsize 0.914} & {\scriptsize 0.962} & {\scriptsize 0.987}   \\   
 {\scriptsize 700} &{\scriptsize 0.973} &{\scriptsize 0.952}  &{\scriptsize 0.974} & {\scriptsize 0.927} & {\scriptsize 0.932}   \\ 

  & & {\scriptsize $\mu =1$}& {\scriptsize $\sigma=0.5$} & \\
 {\scriptsize 50} &{\scriptsize 0.882} &{\scriptsize 0.852} &{\scriptsize 0.823} & {\scriptsize 0.734} & {\scriptsize 0.872}   \\ 
 {\scriptsize 200} &{\scriptsize 0.963} &{\scriptsize 0.880} &{\scriptsize 0.780} & {\scriptsize 0.925} & {\scriptsize  0.967}  \\  
 {\scriptsize 700} &{\scriptsize 0.836} &{\scriptsize 0.925}  &{\scriptsize 0.743} & {\scriptsize 0.943} & {\scriptsize 0.780}  \\ 

 & & {\scriptsize $\mu =1$}& {\scriptsize $\sigma=.1$} & \\ 
 {\scriptsize 50} &{\scriptsize 0.230} &{\scriptsize 0.770} &{\scriptsize 0.611} & {\scriptsize 0.587} & {\scriptsize 0.554}  \\  
 {\scriptsize 200} &{\scriptsize 0.101} &{\scriptsize 0.912} &{\scriptsize 0.686} & {\scriptsize 0.769} & {\scriptsize 0.687} \\  
 {\scriptsize 700} &{\scriptsize 0.051} &{\scriptsize 0.949}  &{\scriptsize 0.727} & {\scriptsize 0.880} & {\scriptsize  0.721} \\ 
 
 & & {\scriptsize $\mu =5$}& {\scriptsize $\sigma=1$} & \\ 
 {\scriptsize 50} &{\scriptsize 0.400} &{\scriptsize 0.810} &{\scriptsize 0.723} & {\scriptsize 0.889} & {\scriptsize 0.457}  \\  
 {\scriptsize 200} &{\scriptsize 0.285} &{\scriptsize 0.945} &{\scriptsize 0.852} & {\scriptsize 0.934} & {\scriptsize 0.579}  \\  
 {\scriptsize 700} &{\scriptsize 0.222} &{\scriptsize 0.963} &{\scriptsize 0.967} & {\scriptsize 0.978} & {\scriptsize 0.789} \\ 
 
  & & {\scriptsize $\mu =5$}&{\scriptsize $\sigma=0.5$} & \\ 
 {\scriptsize 50} &{\scriptsize 0.2390} &{\scriptsize 0.852} &{\scriptsize 0.712} & {\scriptsize 0.845} & {\scriptsize 0.831}  \\  
 {\scriptsize 200} &{\scriptsize 0.1390} &{\scriptsize 0.926} &{\scriptsize 0.897} & {\scriptsize 0.915} & {\scriptsize 0.805} \\  
 {\scriptsize 700} &{\scriptsize 0.0817} &{\scriptsize 0.956}  &{\scriptsize 0.945} & {\scriptsize 0.921} & {\scriptsize 0.878}   \\ 
 
 & & {\scriptsize $\mu =5$}& {\scriptsize $\sigma=0.1$} & \\
 {\scriptsize 50} &{\scriptsize 0.1360} &{\scriptsize 0.588}  &{\scriptsize 0.702} & {\scriptsize 0.645} & {\scriptsize 0.613} \\  
 {\scriptsize 200} &{\scriptsize 0.0523} &{\scriptsize 0.458} &{\scriptsize 0.764} & {\scriptsize 0.758} & {\scriptsize 0.802} \\ 
 {\scriptsize 700} &{\scriptsize 0.0205} &{\scriptsize 0.341} &{\scriptsize 0.861} & {\scriptsize 0.655} & {\scriptsize 0.841} \\ 
 \hline 
 
 \end{tabular}  
\end{small}
\end{center}
\end{table}

\begin{table}
\begin{center}
\begin{small}
\caption{$E\left( \widehat{h} \right) $ for normal mixture $f(x)=0.5 \phi (x)+0.5\phi_{\sigma}(x-\mu)$ \label{tab-id_table2}}
\begin{tabular}{cccccccccc}
  \hline 
  {\small $n$} & {\small $\widehat{h}_{NR}$} & $\widehat{h}_{LSCV}$ & {\small $\widehat{h}_{{\cal D}_{1.1}CV}$} & {\small $\widehat{h}_{{\cal D}_{1.5}CV}$} & {\small $\widehat{h}_{{\cal D}_{1.9}CV}$} & $h_{MISE}$ \\
 \hline 
  & & &  {\scriptsize $\mu =0$}&{\scriptsize $\sigma=1$} & & \\
 {\scriptsize 50} & {\scriptsize 0.464} & {\scriptsize 0.528} & {\scriptsize 0.530} & {\scriptsize 0.520} & {\scriptsize 0.323} & {\scriptsize \textbf{0.347}}   \\ 
 {\scriptsize 200} & {\scriptsize \textbf{0.362}} & {\scriptsize \textbf{0.393}} & {\scriptsize \textbf{0.399}} & {\scriptsize \textbf{0.383}} & {\scriptsize \textbf{0.321}} & {\scriptsize \textbf{0.328}}   \\   
{\scriptsize 700} & {\scriptsize \textbf{0.287}} & {\scriptsize \textbf{0.302}} & {\scriptsize \textbf{0.310}} & {\scriptsize \textbf{0.293}} & {\scriptsize \textbf{0.308}}& {\scriptsize \textbf{0.309}} \\ 
 
  & & &  {\scriptsize $\mu =0$ }& {\scriptsize $\sigma=0.5$ }  &  \\
 {\scriptsize 50 } & {\scriptsize \textbf{0.330}} & {\scriptsize \textbf{0.397}} &{\scriptsize \textbf{0.425}} &{\scriptsize \textbf{0.343}} &{\scriptsize \textbf{0.223}}& {\scriptsize \textbf{0.286}} \\ 
{\scriptsize 200} &{\scriptsize \textbf{0.248}} &{\scriptsize \textbf{0.267}} &{\scriptsize \textbf{0.312}}&{\scriptsize \textbf{0.248}}& {\scriptsize \textbf{0.193}}& {\scriptsize \textbf{0.280}} \\  
{\scriptsize 700} &{\scriptsize \textbf{0.196}} &{\scriptsize \textbf{0.197}} &{\scriptsize \textbf{0.242}} &{\scriptsize \textbf{0.200}} &{\scriptsize \textbf{0.186}} & {\scriptsize \textbf{0.244}} \\ 

  & & & {\scriptsize $\mu =0$}& {\scriptsize $\sigma=0.1$ }& \\
{\scriptsize 50} & {\scriptsize \textbf{0.134}} &{\scriptsize \textbf{0.104}} &{\scriptsize \textbf{0.358}} &{\scriptsize \textbf{0.098}} &{\scriptsize 0.510} & {\scriptsize \textbf{0.041}} \\ 
 {\scriptsize 200} &{\scriptsize \textbf{0.087}} &{\scriptsize \textbf{0.060}} & {\scriptsize \textbf{0.027}} &{\scriptsize \textbf{0.087}} &{\scriptsize 0.485} & {\scriptsize \textbf{0.038}} \\ 
{\scriptsize 700} &{\scriptsize \textbf{0.068}} & {\scriptsize \textbf{0.043}} & {\scriptsize \textbf{0.219}} &{\scriptsize \textbf{0.057}} &{\scriptsize 0.421}& {\scriptsize \textbf{0.0370}}  \\  
  
  & & & {\scriptsize $\mu =1$}& {\scriptsize $\sigma=1$} & \\ 
{\scriptsize 50 }& {\scriptsize 0.520 } &{\scriptsize 0.590 } &{\scriptsize 0.592 } &{\scriptsize 0.588 }&{\scriptsize 0.429} &{\scriptsize 0.426} \\ 
 {\scriptsize 200 }&{\scriptsize 0.404}&{\scriptsize 0.437} &{\scriptsize 0.444}&{\scriptsize 0.434}&{\scriptsize \textbf{0.395}}& {\scriptsize 0.423}   \\
 {\scriptsize 700 }&{\scriptsize \textbf{0.316}}&{\scriptsize 
 \textbf{0.336}} &{\scriptsize 0.344} & {\scriptsize\textbf{0.333}}&{\scriptsize \textbf{0.354}} & {\scriptsize 
 \textbf{0.345}} \\  

  & & & {\scriptsize $\mu =1$}&{\scriptsize  $\sigma=0.5$ }& \\
{\scriptsize 50 }&{\scriptsize 0.401 }&{\scriptsize \textbf{0.430}} &{\scriptsize \textbf{0.479}}&{\scriptsize 0.373}&{\scriptsize \textbf{0.326}}& {\scriptsize \textbf{0.342}} \\ 
{\scriptsize 200 }&{\scriptsize \textbf{0.320}}&{\scriptsize \textbf{0.298}} &{\scriptsize \textbf{0.373}} &{\scriptsize \textbf{0.265}} &{\scriptsize \textbf{0.280}}& {\scriptsize \textbf{0.282}}  \\  
 {\scriptsize 700} & {\scriptsize \textbf{0.254}} & {\scriptsize \textbf{0.214}} & {\scriptsize \textbf{0.287}} & {\scriptsize \textbf{0.212}} &{\scriptsize \textbf{0.233}}& {\scriptsize \textbf{0.239}} \\ 

 & & & {\scriptsize $\mu =1$}&{\scriptsize $\sigma=.1$} & \\ 
{\scriptsize 50} &{\scriptsize \textbf{0.366}} &{\scriptsize \textbf{0.103}} &{\scriptsize \textbf{0.464}} &{\scriptsize \textbf{0.203}} &{\scriptsize \textbf{0.0422}}& {\scriptsize \textbf{0.0451}}  \\  
{\scriptsize 200} &{\scriptsize \textbf{0.276}} &{\scriptsize \textbf{0.061}}&{\scriptsize \textbf{0.342}}&{\scriptsize \textbf{0.053}} &{\scriptsize \textbf{0.0380}} & {\scriptsize \textbf{0.0380}} \\  
 {\scriptsize 700} &{\scriptsize \textbf{0.221}}&{\scriptsize \textbf{0.0428}} & {\scriptsize \textbf{0.267}} &{\scriptsize \textbf{0.0426}}&{\scriptsize \textbf{0.0343}} & {\scriptsize \textbf{0.0314}} \\ 
 
 & & & {\scriptsize $\mu =5$}& {\scriptsize $\sigma=1$} & \\ 
{\scriptsize 50} &{\scriptsize 1.290} &{\scriptsize 0.770} &{\scriptsize 1.400} &{\scriptsize 0.608} &{\scriptsize 0.420} & {\scriptsize 0.475}  \\  
{\scriptsize 200} &{\scriptsize 0.989} &{\scriptsize 0.477} &{\scriptsize 0.1.070} &{\scriptsize 0.441} &{\scriptsize \textbf{0.330}} & {\scriptsize 0.470} \\  
 {\scriptsize 700} &{\scriptsize 0.768}&{\scriptsize \textbf{0.353}} &{\scriptsize \textbf{0.829}} &{\scriptsize \textbf{0.336}}  &{\scriptsize 0.442} &{\scriptsize \textbf{0.272}}  \\ 
 
  & & & {\scriptsize  $\mu =5$}& {\scriptsize $\sigma=0.5$} & \\ 
{\scriptsize  50} & {\scriptsize 1.270} & {\scriptsize 0.468 }& {\scriptsize 1.370}&{\scriptsize 0.369}&{\scriptsize 0.310}& {\scriptsize \textbf{0.295}} \\
{\scriptsize 200 } & {\scriptsize 0.961}& {\scriptsize \textbf{0.297}} & {\scriptsize \textbf{1.040}} & {\scriptsize \textbf{0.262}}&{\scriptsize \textbf{0.210}} & {\scriptsize \textbf{0.286}} \\  
 {\scriptsize 700} & {\scriptsize 0.750} & {\scriptsize \textbf{0.208}}& {\scriptsize \textbf{0.810}}& {\scriptsize \textbf{0.197}}& {\scriptsize \textbf{0.209}}& {\scriptsize \textbf{0.270}}  \\ 
 
 & & & {\scriptsize $\mu =5$}& {\scriptsize $\sigma=0.1$ }& \\
 {\scriptsize 50} & {\scriptsize 1.270} &{\scriptsize 
 \textbf{0.0982}} & {\scriptsize 1.370} & {\scriptsize 
 \textbf{0.0745}} & {\scriptsize \textbf{0.045}}& {\scriptsize  \textbf{0.0415}} \\  
 {\scriptsize 200} & {\scriptsize 0.955}& {\scriptsize \textbf{0.061}} & {\scriptsize \textbf{1.030}}& {\scriptsize \textbf{0.053}} &{\scriptsize \textbf{0.040}}& {\scriptsize \textbf{0.0385}} \\ 
 {\scriptsize 700} & {\scriptsize 0.745} & {\scriptsize \textbf{0.0424}} & {\scriptsize \textbf{0.804}} & {\scriptsize \textbf{0.040}} & {\scriptsize \textbf{0.039}} & {\scriptsize \textbf{0.0339}}  \\ 
 \hline 
 
 \end{tabular}  
\end{small}
\end{center}
\end{table}

\begin{table}
\begin{center}
\begin{small}
\caption{$E \vert \widehat{h}/h_{MISE}-1 \vert $ for normal mixture $f(x)=0.5 \phi (x)+0.5\phi_{\sigma}(x-\mu)$ \label{tab-id_table3}}
\begin{tabular}{ccccccccc}
  \hline 
  {\small $n$} &  {\small $\widehat{h}_{NR}$} & {\small $\widehat{h}_{LSCV}$} & {\small $\widehat{h}_{{\cal D}_{1.1}CV}$}& {\small $\widehat{h}_{{\cal D}_{1.5}CV}$} & {\small $\widehat{h}_{{\cal D}_{1.9}CV}$} \\
 \hline 
  & &{\scriptsize $\mu =0$}&{\scriptsize $\sigma=1$} & \\
 {\scriptsize 50} &{\scriptsize 0.124} &{\scriptsize 0.072} & {\scriptsize 0.077} &{\scriptsize 0.0874} &{\scriptsize 0.379}   \\ 
{\scriptsize 200} &{\scriptsize 0.0785} &{\scriptsize 0.0829} & {\scriptsize 0.1050} &{\scriptsize 0.0578} &{\scriptsize 0.1620} \\   
{\scriptsize 700} & {\scriptsize 0.0396} &{\scriptsize 0.0717} & {\scriptsize 0.0572} &{\scriptsize  0.0436} &{\scriptsize 0.0509}  \\ 
 
  & & {\scriptsize $\mu =0$}& {\scriptsize $\sigma=0.5$}  & \\
{\scriptsize 50 }&{\scriptsize \textbf{0.1370}} &{\scriptsize \textbf{0.1510}} &{\scriptsize \textbf{0.1670}} &{\scriptsize \textbf{0.1510}} &{\scriptsize 0.4560} \\ 
{\scriptsize 200} &{\scriptsize \textbf{0.0655}} &{\scriptsize \textbf{0.1360}} &{\scriptsize \textbf{0.0882}} &{\scriptsize \textbf{0.1010}} &{\scriptsize \textbf{0.2490}}     \\  
{\scriptsize 700} &{\scriptsize \textbf{0.0559}} &{\scriptsize \textbf{0.0729}} &{\scriptsize \textbf{0.0537}} &{\scriptsize \textbf{0.0818}} &{\scriptsize \textbf{0.0104}}     \\ 

  & &  {\scriptsize$\mu =0$}& {\scriptsize$\sigma=0.1$} & \\
 {\scriptsize 50} &{\scriptsize 0.772} &{\scriptsize \textbf{0.2530}} & {\scriptsize 0.3000} &{\scriptsize 0.3990} &{\scriptsize 0.4410}  \\ 
{\scriptsize 200} &{\scriptsize 0.674} &{\scriptsize \textbf{0.1210}} &{\scriptsize \textbf{0.1250}} &{\scriptsize \textbf{0.1520}} &{\scriptsize \textbf{0.2070}}   \\ 
{\scriptsize 700} &{\scriptsize 0.679} &{\scriptsize \textbf{0.0772}} &{\scriptsize \textbf{0.0506}} &{\scriptsize \textbf{0.0726}} &{\scriptsize \textbf{0.185}}  \\  
  
  & &  {\scriptsize$\mu =1$}& {\scriptsize$\sigma=1$} & \\ 
 {\scriptsize 50} &{\scriptsize 0.1430} &{\scriptsize 0.1040} &{\scriptsize 0.1630} &{\scriptsize 0.0748} &{\scriptsize 0.398} \\ 
{\scriptsize 200} &{\scriptsize 0.0774} &{\scriptsize  0.0800} &{\scriptsize 0.0931} &{\scriptsize  0.0501} &{\scriptsize 0.184}  \\  
 {\scriptsize 700 }&{\scriptsize 0.0483} &{\scriptsize 0.0626} &{\scriptsize 0.0600} &{\scriptsize 0.0361} &{\scriptsize 0.037} \\  

  & &  {\scriptsize $\mu =1$}& {\scriptsize$\sigma=0.5$} & \\
{\scriptsize 50} &{\scriptsize 0.172} &{\scriptsize 0.1930} &{\scriptsize 0.1400} &{\scriptsize 0.2260} &{\scriptsize 0.4560} \\ 
{\scriptsize 200} &{\scriptsize 0.236} &{\scriptsize 0.1530} &{\scriptsize 0.0899} &{\scriptsize 0.1460} &{\scriptsize 0.121} \\  
{\scriptsize 700} &{\scriptsize 0.285} &{\scriptsize 0.0989} &{\scriptsize 0.0506} &{\scriptsize 0.0794} &{\scriptsize 0.119} \\ 

 & &  {\scriptsize $\mu =1$}&{\scriptsize $\sigma=.1$} & \\ 
{\scriptsize 50} &{\scriptsize 3.67} &{\scriptsize 0.2620} &{\scriptsize 1.380} &{\scriptsize 0.3980} \\  
{\scriptsize 200} &{\scriptsize 4.29} &{\scriptsize 0.1250} &{\scriptsize 0.986} &{\scriptsize 0.1720} &{\scriptsize 0.353} \\  
{\scriptsize 700} &{\scriptsize 4.58} &{\scriptsize 0.0838} &{\scriptsize 0.652} &{\scriptsize 0.0878} &{\scriptsize  0.137} \\ 
 
 & &  {\scriptsize $\mu =5$}&{\scriptsize $\sigma=1$} & \\ 
{\scriptsize 50} &{\scriptsize 1.14} &{\scriptsize 0.1450} &{\scriptsize 0.2390} &{\scriptsize 0.2430} &{\scriptsize 0.4580}  \\  
{\scriptsize 200} &{\scriptsize 1.23} &{\scriptsize 0.0815} &{\scriptsize 0.1210} &{\scriptsize 0.0899} &{\scriptsize 0.2530}   \\  
{\scriptsize 700} &{\scriptsize 1.29} &{\scriptsize 0.0686} &{\scriptsize 0.0745} &{\scriptsize 0.0540} &{\scriptsize 0.0203}  \\ 
 
  & &  {\scriptsize$\mu =5$}& {\scriptsize $\sigma=0.5$} & \\ 
 {\scriptsize 50} &  {\scriptsize 2.40} &  {\scriptsize 0.1860} & {\scriptsize 0.600} & {\scriptsize 0.2510} & {\scriptsize 0.4340} \\  
 {\scriptsize 200} & {\scriptsize 2.68} & {\scriptsize 0.1180} & {\scriptsize 0.444} & {\scriptsize 0.1440} & {\scriptsize  0.2020} \\  
 {\scriptsize 700} & {\scriptsize 2.80} & {\scriptsize 0.0743} & {\scriptsize 0.296} & {\scriptsize 0.0804} & {\scriptsize 0.0597}  \\ 
 
 & &  {\scriptsize $\mu =5$}&  {\scriptsize$\sigma=0.1$} & \\
 {\scriptsize 50} & {\scriptsize 15.7} & {\scriptsize 0.884} & {\scriptsize 5.95} & {\scriptsize 0.3980} & {\scriptsize 0.4810}   \\  
 {\scriptsize 200} & {\scriptsize 17.1} & {\scriptsize 1.021} & {\scriptsize 4.51} & {\scriptsize 0.1570} & {\scriptsize 0.2630}  \\ 
 {\scriptsize 700} & {\scriptsize 17.7} & {\scriptsize 1.104} & {\scriptsize 3.34} & {\scriptsize 0.0656} & {\scriptsize 0.0178} \\ 
 \hline 
 
 \end{tabular}  
\end{small}
\end{center}
\end{table}

\begin{figure}
\begin{center}
\label{re}
\includegraphics[scale=0.47]{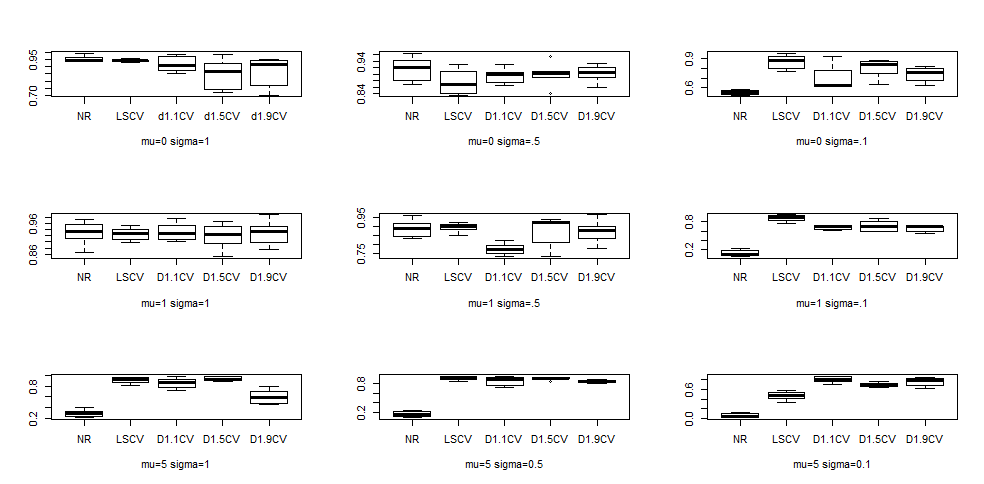}
\end{center}
\caption{Boxplots of the relative values Re forthe bandwidth selectors forestimation of the densities $\mu =0, 1, 5$ and $\sigma =1, 0.5, 0.1$. The sample size varies from $100$ to $2000$. }
\end{figure}

Figure 1 compare, for densities with $(\mu =0, 1, 5$ and $\sigma=1, .5, .1 )$, the results of the five bandwidth selection $NR$, $LSCV$ and ${\cal D}_{\beta}$ (discussed in Section \ref{choi}), relatively to the results obtained by using the $MISE$ optimal bandwidth ($h_{MISE}$). These figures present boxplots of the ratio $RE(\widehat{h})=MISE(\widehat{f}_{\widehat{h}_{MISE}})/MISE(\widehat{f}_{\widehat{h}})$ for each bandwidths $\widehat{h}_{NR}$, $\widehat{h}_{LSCV}$ and $\widehat{h}_{{\cal D}_{\beta}}$ (with $\beta =1.1, 1.5$ and $1.9$). We see  the $LSCV$ and ${\cal D}_{\beta}$ (with $\beta =1.5$) methods gave overall the bests ratios across all simulations, and that this ratio was rather large in general.

\section{Illustration with real data}
Two examples are provided to demonstrate the performance of kernel density estimation with different bandwidths, where the Gaussian kernel is used. All of them are classical examples of unimodal and bimodal distributions, respectively.\\

\textbf{Example 1}
\\
The first data set comprises the CO2 per capita in the year of 2014. The data set can be downloaded from the world bank website.\\
Figure 2 shows the estimated density of CO2 per capita in the year of 2008, , we using bandwidths $\widehat{h}_{NR}=1.38$, $\widehat{h}_{LSCV}=0.439$, $\widehat{h}_{1.5}=0.832$, $\widehat{h}_{1.1}=0.932$ and $\widehat{h}_{1.9}=0.542$ \\
The data set that the estimated density that was computed with the $\widehat{h}_{LSCV}=0.439$ and  $\widehat{h}_{1.9}$  bandwidths captures the peak that characterizes the mode, while the estimated density with the bandwidths that $\widehat{h}_{NR}$, $\widehat{h}_{1.5}$ and $\widehat{h}_{1.1}$ smoothes out this peak. 

This happens because the outliers at the tail of the distribution contribute to  $\widehat{h}_{NR}$,  $\widehat{h}_{1.5}$ and  $\widehat{h}_{1.1}$ be larger than the than other bandwidths..
\begin{figure}[h]
\begin{center}
\label{exponentiel}
\includegraphics[scale=0.6]{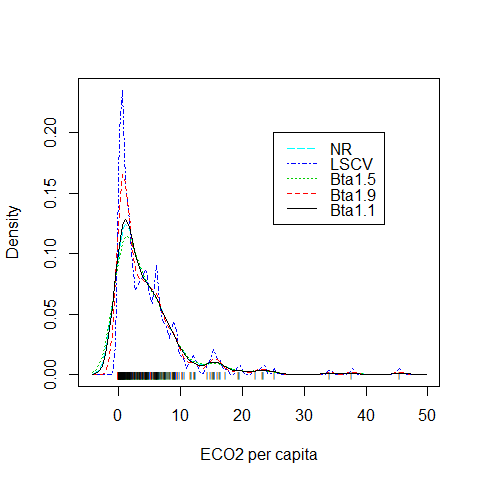}
\end{center}
\caption{Estimated density of CO2 per capita in 2008 using the different bandwidths.  ${\cal D}_{1.1}$
(solid line);  ${\cal D}_{1.9}$ (dashed line);  ${\cal D}_{1.5}$ (dotted line); $LSCV$, least squares cross-validation (dotdash line) and $NR$, normal reference (longdash line). }
\end{figure}
\\

\textbf{Example 2}
we use the time between eruptions set for the Old Faithful geyser in Yellowstone National Park, Wyoming, USA (107 sample data, source: Silvermanl [21]).
\\
Figure 3 plot the data points and the kernel density 
estimates for old faithful geyser data, we using bandwidths  
$\widehat{h}_{NR}=0.442$, $\widehat{h}_{LSCV}=0.162$, $
\widehat{h}_{1.5}=0.176$, $\widehat{h}_{1.1}
=0.281$ and $\widehat{h}_{1.9}=0.210$.\\
An important point to note that the density curve for eruption length is similar to bimodal normal density (normal mixture). From our example 2 we see that the $h_{NB}$ is always larger than  the others bandwidths, he heavily oversmoothes its kernel density curve, underestimating the two peaks of the curve but overestimating the valley between them. About $h_{LSCV}$,  $\widehat{h}_{1.5}$ and  $\widehat{h}_{1.9}$ seems to undersmooth the curve too much, overestimating the two peaks but underestimating for the valley.
However $\widehat{h}_{1.1}$ is proper bandwidth for their density estimate to be able to capture the feature of the true density curve.

\begin{figure}
\begin{center}
\label{exponentiel}
\includegraphics[scale=0.6]{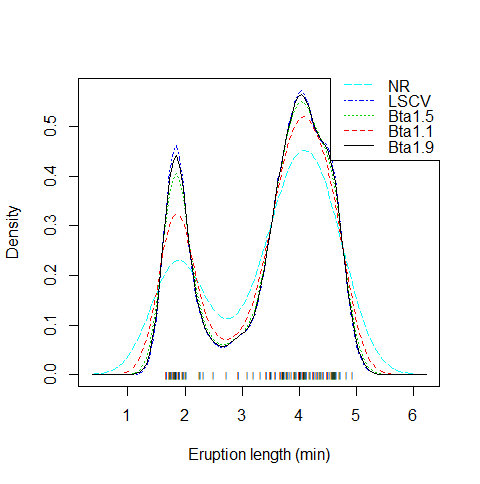}
\end{center}
\caption{Estimated density of Old Faithful geyser using the different bandwidths.  ${\cal D}_{1.1}$
(solid line);  ${\cal D}_{1.9}$ (dashed line);  ${\cal D}_{1.5}$ (dotted line); $LSCV$, least squares cross-validation (dotdash line) and $NR$, normal reference (longdash line). }
\end{figure}

%\textbf{Example 3} \\
%We use the measurements of the directions taken by 76 turtles after being removed from their home territory carried out by Gould (1957) and data used later by Stephens(1969), Fisher (1989), Mardia and Jupp (2000) and Rao and SenGupta (2001).

%%%%%%%%%%%%%%%%%%%%%%%%%%%%%%%%%%%%%%%%%%%%%%%%%%%%%%%%%%%5
\section{Conclusions}
This paper proposed the method for bandwidth selection of  bias reduction kernel density estimator, given in (2). A various bandwidth selection strategies have been proposed such as normal reference $h_{NR}$, least squares cross-validation $h_{LSCV}$ and $h_{\beta}$ for minimizing criterion $\beta$-divergence (with $\beta = 1.5,  1.1$ and $1.9$).\\
The normal reference bandwidth $h_{NR}$ method is a simple and quick selector, but limited the practical use  ., since they are restricted to situations where a pre-specified family of densities is correctly selected. \\
The $LSCV$ do not provide a smooth density estimation, although asymptotically optimal, the finite sample behavior of $h_{LSCV}$ is disappointing for its variability and undersmoothing.\\
We have attempted to evaluate choice the optimal bandwidth
$h_{LSCV}$ and $h_{NR}$, using $\beta$-divergence. Compared to traditional bandwidth selection methods designed for kernel density estimation, our proposed ${\cal D}_{\beta}$ bandwidth selection method is always one of the best for having large $RE(\widehat{h})$ and small $\mathbb{E}(\widehat{h}/\widehat{h}_{MISE}-1) $.\\
Simulation studies showed that our proposed optimal bandwidth ${\cal D}_{\beta}$ method designed for kernel density estimation  adapts to different situations, and out-performs other bandwidths.
we conclude that the choice of the bandwidth based on the real data is consistent with the one based on simulations which is the ${\cal D}_{\beta}$ $\beta = 1.1$ and $1.5$ ) method gives us a smoother density estimation.

\end{document}